\documentclass[showpacs,preprintnumbers,amsmath,amssymb]{revtex4}

\usepackage{graphicx}
\usepackage{dcolumn}
\usepackage{bm}

\renewcommand{\baselinestretch}{2}
\begin{document}

\preprint{APS/123-QED}

\title{M\"{o}bius and twisted graphene nanoribbons: \\ stability, geometry and electronic properties}

\author{E. W. S. Caetano}
\email{ewcaetano@gmail.com}
\affiliation{Centro Federal de Educa\c{c}\~ao Tecnol\'ogica do Cear\'a, Avenida 13 de Maio, 2081, Benfica, 60040-531 Fortaleza, Cear\'a, Brazil}

\author{V. N. Freire}
\author{S. G. dos Santos}
\affiliation{Departamento de F\'{\i}sica, Universidade Federal do Cear\'a, Centro de Ci\^encias, Caixa Postal 6030, Campus do Pici, 60455-760, Fortaleza, Cear\'a, Brazil}

\author{D. S. Galv\~{a}o}
\author{F. Sato}
\affiliation{Instituto de F\'{\i}sica 'Gleb Wataghin', Universidade Estadual de Campinas, Unicamp, 13083-970, Campinas, S\~{a}o Paulo, Brazil}

\date{\today}

\begin{abstract}
Results of classical force field geometry optimizations for twisted graphene nanoribbons with a number of twists $N_t$ varying from 0 to 7 (the case $N_t$=1 corresponds to a half-twist M\"{o}bius nanoribbon) are presented in this work. Their structural stability was investigated using the Brenner reactive force field. The best classical molecular geometries were used as input for semiempirical calculations, from which the electronic properties (energy levels, HOMO, LUMO orbitals) were computed for each structure. CI wavefunctions were also calculated in the complete active space framework taking into account eigenstates from HOMO-4 to LUMO+4, as well as the oscillator strengths corresponding to the first optical transitions in the UV-VIS range. The lowest energy molecules were found less symmetric than initial configurations, and the HOMO-LUMO energy gaps are larger than the value found for the nanographene used to build them due to electronic localization effects created by the twisting. A high number of twists leads to a sharp increase of the HOMO $\rightarrow$ LUMO transition energy. We suggest that some twisted nanoribbons could form crystals stabilized by dipolar interactions.
\end{abstract}

%\keywords{Graphene \sep nanoribbons \sep M\"{o}bius \sep classical force field simulations \sep electronic structure}

\maketitle

\section{Introduction}

In 1858, the German mathematician August Ferdinand M\"{o}bius investigated the properties of a surface obtained by gluing the extremities of a twisted band. More than one century later, molecules with the shape of a half-twist M\"{o}bius band were synthesized for the first time \cite{r1}. In 1964, Heilbronner \cite{r2}  proposed the definition of a M\"{o}bius aromaticity for cyclic molecules, foreseeing that M\"{o}bius aromatic hydrocarbons with $4n$ p-electrons should be stable, in contrast with  the H\"{u}ckel rule prediction on the instability of antiaromatic hyrdrocarbons with $4n$ p-electrons. According to Heilbronner calculations, the twist changes the electronic structure of a molecular ring to a closed shell configuration. It was also demonstrated that the topological change to a half-twist M\"{o}bius band leads to a simple and general relationship between the electronic energy eigenvalues of the corresponding molecules \cite{r3}, giving an elegant description of pure p-like systems over a nonorientable (one-sided) surface.

In a more recent work, Mauksch \textit{et al.} \cite{r4} presented a computational reinterpretation of experimental data showing that (CH)$_9^+$ could be a M\"{o}bius aromatic cyclic annulene with $4n$ $\pi$ electrons. Electronic properties of ring compounds were discussed theoretically in connection with M\"{o}bius aromatic properties by some authors \cite{r5,r6,r7,r8,r9}. Mart\'{\i}n-Santamaria and Rzepa \cite{r5} presented an analysis of the main features of M\"{o}bius annulenes considering the $\pi$-molecular orbital correlation between the planar H\"{u}ckel configuration and the twisted C$_2$ symmetric M\"{o}bius system. Their calculations were performed within both the frameworks of the semiempirical SCF-MO method and the restricted B3LYP/6-31G(d) level of \textit{ab initio} Density Functional Theory (DFT). They observed that the form of the calculated AM1 and B3LYP/6-31G(d) orbitals was very similar, with minor variations in the relative order of the occupied orbitals. They concluded that when ring C-H bonds are present in a cyclic annulene the M\"{o}bius orbitals can interact significantly with the $\sigma$ framework, creating a "ghost" $\pi$-like H\"{u}ckel orbital.  In other paper \cite{r6}, Mart\'{\i}n-Santamaria and Rzepa obtained AM1 and \textit{ab initio} geometries of M\"{o}bius bands formed by imposing one, two or three twists to cyclacenes of various lengths, showing that there is a localisation of the twist over 2-4 benzene rings, and reporting the geometries, highest occupied molecular orbitals, electrostatic potential and charge distributions for neutral, 6+ and 6- forms of these molecules.  Rzepa \cite{r7} carried out B3LYP and KMLYP/6-31G(d) calculations that predicted a double-helical and chiral conformation of [14]annulene with the topological properties of a double-twist M\"{o}bius band. In another work \cite{r9}, the same author summarized the evidence to support Heilbronner's prediction of a counterpart to planar H\"{u}ckel aromatic rings with M\"{o}bius topology and proposes higher order twisted systems (paradromic rings or Listing rings) as interesting candidates for future study.

The synthesis of a stable M\"{o}bius aromatic hydrocarbon was obtained for the first time in 2003 by Ajami \textit{et al.} \cite{r10} who combined a 'normal' aromatic structure (like benzene, with trigonal planar sp$^2$-hybridized atoms) and a 'belt-like' aromatic structure (like the surface of a carbon nanotube, with pyramidalized sp$^2$ atoms), creating a M\"{o}bius compound stabilized by an extended $\pi$-system. X-ray characterization of crystals grown from the synthesized M\"{o}bius hydrocarbons showed good agreement with DFT calculations. C$_2$ symmetric M\"{o}bius compounds lead to red crystals, whereas the C$_s$ symmetric H\"{u}ckel counterpart compound is colourless. Observed trends in bond-length equalization and stabilization energy point to a moderately aromatic character of the M\"{o}bius structure, whereas the corresponding H\"{u}ckel structure is non-aromatic. Further M\"{o}bius isomers were presented by Ajami \textit{et al.} in 2006 \cite{r11} with details of their preparation as stable compounds and an investigation on their energy, geometry and magnetic parameters. We note also that Starostin and Van Der Heijden \cite{r11a} have used the invariant variational bicomplex formalism to find the first equilibrium equations for a wide strip subjected to large deformations, which can be useful to understand the relationship between geometry and physical properties of nanometric M\"{o}bius structures. A review of the most recent developments on the design of M\"{o}bius molecules can be found in reference \cite{r12}.

The aim of this work is to investigate theoretically the structural stability, optimal geometry, and the electronic and optical properties of a class of twisted bands built from graphene nanoribbons, including the case of a half-twist M\"{o}bius graphene strip. Indeed, graphene is a novel material with very interesting and promising properties, such as thermal conductivity, hardness, and extraordinary electronic properties, like being a zero gap semiconductor with a Dirac-like energy spectrum close to the Fermi level (or, in other words, electrons in this region have zero effective mass and are analogous to photons in Einstein's theory of relativity) \cite{r13,r14,r15,r16,r17,r18,r19,r20,r21}. Zigzag nanoribbons from graphene have been investigated using first principles calculations and it was shown that they can carry a spin current \cite{r22}. Under M\"{o}bius boundary conditions, these nanoribbons have a magnetic domain wall whose width depends on the Coulomb interaction \cite{r23,r24}. A first principles study on the spin states of zigzag graphene nanoribbons has found a triplet ground state for a M\"{o}bius cyclacene and an open-shell singlet ground state for the corresponding two-edge counterpart. Wider nanoribbons with M\"{o}bius topology exhibit an increasing total magnetization with the ribbon length \cite{r25}.

\section{Computational Details}

A single rectangular nanographene (see Figure 1(a)) with armchair and zigzag sides was used for all calculations we have carried out. We can describe it by using two integer parameters: its length ($L$) and width ($W$). The length $L$ is obtained by counting the number of C-C dimers along (and parallel to) the armchair edge. The width $W$, on the other hand, is given by counting the number of C-C dimers orthogonal to the zigzag border. Thus our nanographene has $L=29$ ($\approx 6.8$ nm) and $W=5$ ($\approx 0.5$ nm), its molecular formula being C$_{150}$H$_{60}$ (we have chosen the values for $L$ and $W$ in such a way to achieve a round number of carbon atoms). Closed nanoribbons can be built by joining the nanographene zigzag ends. If we twist the nanographene along its length by an angle of 180$^0$ and join the zigzag extremities, we obtain a half-twist M\"{o}bius nanoribbon (Figure 1(c)). If we twist the nanographene by an angle of N$_t \times 180^0$ and join the zigzag extremities, we build a $N_t$-twisted nanoribbon (Figure 1(d)-(i)). In this work, we have chosen structures with $N_t$ varying from 0 to 7. Dangling bonds were passivated with hydrogen atoms during the simulations.

We started out by performing classical geometry optimizations for each nanoribbon using the Universal force field, available in the Forcite code. The following convergence tolerances were adopted: $2 \times 10^{-5}$ kcal/mol (total energy variation), 0.01 kcal/mol/nm (maximum force per atom), and $10^{-6}$ nm (maximum atomic displacement). An algorithm using a cascade of the steepest descent, adjusted basis set Newton-Raphson and quasi-Newton methods was selected. Non-bond energies (van der Waals and electrostatic) were calculated using an atom based summation method with a cutoff distance of 6 nm. After this optimization, an annealing simulation was carried out to search the conformational space for lowest energy structures through cycles of increasing-decreasing temperatures of a classical dynamics in order to prevent conformations that correspond to local energy minima with higher energies. For the annealing, we have use the following parameters: NVE ensemble, time step of 1 fs, 100 annealing cycles, initial (final) temperature of 300 K (800 K), 50 heating ramps per cycle and 100 dynamic steps per ramp. After each anneal cycle, the lowest energy structure was optimized following the same criteria of the first geometry optimization. Aiming to evaluate the structural stability of the twisted nanoribbons, a classical molecular dynamics using the Brenner reactive forcefield \cite{brenner1,brenner2} and the GULP code was also performed starting from the optimized geometries and using a NVE ensemble with a temperature of 500 K, time step of 1 fs, equilibration time of 500 ps, and production time of 500 ps.

The electronic structure was simulated for all nanoribbons using the AM1 semiempirical Hamiltonian. Previous reports in the literature \cite{am1a,am1b,am1c,am1d} suggest that both AM1 and first principles Hartree-Fock results are really comparable for both carbon and silicon cages and provide firm support to the applicability of the AM1 method to study all-silicon and all-carbon clusters. Thus we think it is appropriate to use the semiempirical AM1 Hamiltonian to study the graphene nanoribbons as well. Wavefunctions were calculated within the re\-strict\-ed Hartree-Fock approximation (ground state in a singlet spin state), and a geometry optimization was carried out starting from the best molecular conformations achieved through the classical annealing. The lowest vibrational frequencies were monitored for negative values to ensure that the semiempirical conformations correspond to local energy minima. Electronic energies and 660 self-consistent molecular orbitals (330 occupied) were then obtained. After the semiempirical geometry optimization, the configuration interaction (CI) wavefunctions for the singlet ground state and the first 63 excited states were calculated using the complete active space method and taking into account the molecular orbitals (MOs) from HOMO-4 to LUMO+4 (10 MOs). From these CI wavefunctions, the energies corresponding to the first optical transitions and their respective oscillator strengths were computed.

\section{Results and discussion}

\subsection{Geometry and stability of the twisted nanoribbons}

In Table 1 we present the space point groups of symmetry for the nanoribbons and the rectangular nanographene used to build them. The nanographene has $D_{2h}$ symmetry (8 symmetry operations), whereas the initial configuration for the 0-nanoribbon is the most symmetrical of all, belonging to the $D_{15h}$ group (60 symmetry operations). The classically optimized 0-nanoribbon (after the annealing) exhibits a much smaller degree of symmetry ($C_2$ point group). The semiempirical optimization, on the other hand, leads to a 0-nanoribbon with symmetry belonging to the $C_{5h}$ point group (10 symmetry operations). Much of this symmetry breaking is due to the insertion of hydrogen atoms to passivate the dangling bonds at the nanoribbon edges, which creates more degrees of freedom for the annealing simulation to explore asymmetrical geometries. Indeed, the symmetry decreases with respect to the initial inputs for all the other nanoribbons, from $N_t=1$ (half-twist M\"{o}bius) to 7, after the annealing and semiempirical simulations. We can say therefore that the twist localization created by such symmetry breaking is not merely a quantum mechanical effect, but a more general mechanical feature of twisted ribbons, as pointed in the paper of Starostin and Van Der Heijden \cite{r11a}. The half-twist M\"{o}bius nanoribbon ($N_t=1$) has an initial $C_2$  point group, but ends at a $C_1$ geometry for both the classical and semiempirical results. Nanoribbons with $N_t=2,4,7$ start with $C_2$ symmetry and end with $C_1$ geometries as well. The $N_t=3,6$ molecules end both at $C_2$ symmetry in the classical and $C_1$ in the semiempirical simulation. Finally, for $N_t=5$, we have an initial atomic structure with $C_5$ symmetry that ends with $C_2$ point group for both classical and semiempirical calculations. In comparison, Mart\'{\i}n-Santamar\'{\i}a and Szepa \cite{r6}, using semiempirical and ab initio methods, have studied single, double and triple twisted M\"{o}bius cyclacenes, and have shown that the optimized geometries have localized twists and reduced symmetry, agreeing with the results we obtained for the twisted armchair nanoribbons. We believe this agreement points to localizing behavior as a general topological feature of M\"{o}bius systems.

In a classical molecular mechanics simulation the total energy of a system is usually decomposed in contributions due to the stretching of bonds, the bending of angles between bonds from the same atom, the torsion of dihedral angles, the existence of out-of-plane interactions, and the non-bond interactions (van der Waals, Coulomb, hydrogen bonds). Table 2 shows the total energy and the contribution from valence and non-bond terms (only van der Waals) for all nanoribbons from $N_t=0$ to 7. We can see that the total energy increases with the number of twists $N_t$. A quadratic fit shows that the total energy $E_{\rm{TOT}}$ as a function of $N_t$ (with maximum error of about 5\%) can be given  by  $E_{\rm{TOT}}(N_t)=311.94-16.82N_t+11.47N_t^2$. The $N_t=7$ strip has a total energy 2.6 times larger than the total energy for a nanoribbon with $N_t=0$. The variation of energy from $N_t=0$ to $N_t=1$ (half-twist M\"{o}bius) is 24.91 kcal/mol, corresponding to a relative increase of about 8.4\%. This total energy increase behavior was also observed by Fowler and Rzepa using a pure H\"{u}ckel treatment for cycles with an arbitrary number of half-twists \cite{arrules}.

Looking at the terms contributing to the total energy, we see that the energy related to the stretching of bonds is practically the same (around 30 kcal/mol) for all nanoribbons from $N_t=0$ to 5, indicating that the change in bond lengths is not remarkable for structures with $N_t<6$. On the other hand, for $N_t=6,7$ we note a more significant increase of the bond energy term, pointing to an increase in the bond strain as the number of torsions crosses a structural threshold. Bond energies contribute with 11\% of the total energy at most (for $N_t=0$) and this figure is smaller for larger values of $N_t$ (reaching about 5\% for $N_t=7$). The energy term related to the bending of covalent bonds, however, is increasingly more important as we switch from 0 to 7-nanoribbons. It is smaller than the bond energy for $N_t<4$, but increases to approximately 12.5\% of the total energy for $N_t=7$.

It is expected that, as the nanoribbons are subjected to more and more twists, the energy related to the torsion of dihedral angles must increase almost linearly with $N_t$. A linear interpolation, however, is not a good approximation for the behavior observed in our calculations. As occurred with the total energy, the torsion energy increase with $N_t$ can be modeled by a parabolic fit according to $E_{\rm{TORSION}}(N_t)=65.9+11.3N_t+6.0N_t^2$ (maximum error of 9\%). The relative contribution of the torsion term to the total energy increases from 21.3\% ($N_t=0$) to 57.3\% ($N_t=7$), becoming dominant (larger than the van der Waals energy) for $N_t>2$. The van der Waals is the only non-bond energy taken into account in the classical calculations carried out in this work, and varies from 61.6\% ($N_t=0$) to approximately 18.5\% ($N_t=7$) relative to the total energy. Its absolute value decreases from $\approx 183$ kcal/mol ($N_t=0$) to $\approx 130$ kcal/mol ($N_t=5$), and the increases to $\approx 137$ kcal/mol ($N_t=6$), and $\approx 142$ kcal/mol ($N_t=7$). Finally, the inversion term is the smallest from all contributions for $N_t<6$, ranging from only 1.79 kcal/mol ($N_t=0$) to 53.18 kcal/mol ($N_t=7$). Notwithstanding, it is the contribution from the  inversion energy to the total energy  that presents the largest relative variation of all terms considered in the calculations, varying from 0.6 \% ($N_t=0$) to 6.9 \% ($N_t=7$).

The chemical stability of all twisted nanoribbons was also investigated through a classical molecular dynamics simulation using the Brenner forcefield, which provides a reactive bond-order potential useful to model organic systems \cite{brenner1,brenner2}. The simulation temperature was set to 500 K in a NVE ensemble, with a time step of 1 fs and a total time of 1 ns. We have not observed any bond dissociation for all nanoribbons studied in this work, which indicates they are chemically stable under the simulated conditions. Beyond that, we mention also the recent work by Rzepa \cite{writh} showing that the stabilization of twisted moieties could be associated with the reduction of local twists of adjacent p$_{\pi}$-p$_{\pi}$ orbitals by the coiling of the central axis of the ring into three dimensions through writhing. Such coiling can indeed be observed in the optimized structures of this work, as shown in Figs. 1 (results from classical simulations) and 3 (quantum semiempirical results).

\subsection{Electronic structure}

As observed for the classical molecular mechanics simulations, the total energy for the geometries optimized using the semiempirical approach shows a quadratic dependence on the number of twists. The heat of formation ($HF$) of the nanoribbons can be given with good accuracy (relative error always smaller than 3\%, being maximal for $N_t=1$, the M\"{o}bius band) by $HF(N_t)=784.58+3.43N_t+15.04N_t^2$ kcal/mol. One sees in Figure 2 the highest occupied molecular orbital (HOMO) and lowest unoccupied molecular orbital (LUMO) eigenenergies for all nanoribbons (and the classically optimized nanographene), as well as the HOMO-LUMO energy gap, the CI energy gap and the energy of the first allowed optical transition. The HOMO and LUMO eigenenergies behave in opposite ways as $N_t$ increase, with the HOMO (LUMO) tending to increase (decrease) slightly with some oscillation as $N_t$ varies from 0 to 5 and decreasing (increasing) sharply for $N_t=6,7$. The HOMO-LUMO gap $E_g$, defined as the difference between the HOMO and LUMO energies, $E_g=E_{\rm{LUMO}}-E_{\rm{HOMO}}$, behaves qualitatively like the LUMO as a function of $N_t$, starting from 3.67 eV for the nanographene rectangle, increasing to 4.65 eV for the 0-nanoribbon and decreasing to 4.58 eV for the 1-nanoribbon (with half-twist M\"{o}bius topology). A comparison with previous calculations within the framework of density functional theory in the generalized gradient approximation (DFT-GGA, PBE exchange-correlation functional) for finite armchair nanographenes \cite{dft} shows that, in the case of a ($W=5, L=12$) molecule, the HOMO-LUMO energy gap is 0.6 eV and, for an infinite 1D-($W=5$) nanographene $E_g, \approx$ 0.3 eV. Considering that the graphene rectangle in our work has $L=29$, we could expect a DFT-GGA energy gap possibly between 0.3 eV and 0.6 eV. To check for this, DFT simulations were carried out (using only the classically optimized geometry of the flat nanographene) and we obtained the same value of $\approx 0.11$ eV for the HOMO-LUMO energy gap using the GGA-PBE, GGA-BLYP, and LDA-VWN exchange-correlation functionals, values smaller than we could expect based in the prediction for the 1D case of reference \cite{dft}. It seems that the HOMO-LUMO gap of the graphene finite ribbon with $W=5$ as a function of $L$ does not decrease monotonically, but oscillates about the value for the $L=\infty$ case. Comparing the DFT estimates with the semiempirical result (3.67 eV), we see that it predicts a much higher HOMO-LUMO gap. We must remember here that, despite including some electron correlation energy via its parameters, the AM1 Hamiltonian produces energy gaps close to the ones obtained through the Hartree-Fock approximation \cite{am1b,nanot}, which does not take into account any correlation effects and, therefore, widens the energy gaps in comparison to experimental values. DFT methods, on the other hand, tend to overestimate electronic correlation, underestimating energy gaps. So we must exercise some caution when analyzing the semiempirical and DFT results. As we will see, the CI calculations include some improvements on estimating electronic correlation energy, predicting energy gaps between the ground state and the first excited state intermediary between the DFT and AM1 estimations (and hopefully more accurate than both) for all nanoribbons. For $N_t=2$, the AM1 HOMO-LUMO energy gap increases in comparison with the $N_t=1$ (half-twist M\"{o}bius) case to 4.66 eV and then decreases for $N_t=3$ down to 4.56 eV. The $N_t=4$ gap is very close to $E_g(N_t=3)$, being 4.55 eV. The smallest gap is obtained for the $N_t=5$ molecule, $E_g(N_t=5)=4.42$ eV. Finally, $E_g(N_t=6)=5.20$ eV and $E_g(N_t=7)=5.45$ eV (the largest ones).

To explain why the HOMO-LUMO energy gap increases sharply for $N_t>5$, we have plotted the HOMO and LUMO orbitals for all nanoribbons and the original nanographene rectangle, as shown in Figure 3. One can see that the nanographene has hydrogen atoms passivating the dangling bonds in its extremities which are absent in the nanoribbons. The nanographene HOMO (LUMO) orbital is delocalized over its length, resembling a metallic quantum state. Indeed, there is a rule for 1D (infinite) nanographenes according to which these nanostructures are metallic if $W=3N-1$, where $N$ is an integer \cite{rule1,rule2,rule3,rule4}. As the finite (0D) nanographene studied here has $W=5=3\times 2-1$, one can see that in this particular case such rule holds. The phase of the HOMO orbital alternates its sign when we move along the nanographene length. We will denote this amplitude pattern $\pi L$. On the other hand, the LUMO orbital has an amplitude that changes its sign as we move across the width of the nanographene, thus we have chosen to indicate this pattern using the $\pi W$ abbreviation. Both patterns were also observed in the previously mentioned calculations using DFT-GGA and DFT-LDA methods to check the nanographene HOMO-LUMO energy gap. Nanoribbons with $N_t$ varying from 0 to 5 have HOMO (LUMO) amplitude phase with $\pi W$ ($\pi L$) aspect, swapping the order of patterns observed for the original nanographene. For $N_t=3,4,5$, electrons occupying the HOMO and LUMO states clearly present a pronounced degree of localization (Figure 3(e,f,g,e',f',g')). For $N_t=6,7$, one can see amplitude phase structures with the same order observed for the nanographene, and for $N_t=7$ electrons in HOMO or LUMO states become strongly localized. So we conclude that the formation of a closed nanoribbon initially inverts the phase patterns of frontier orbitals observed for the armchair nanographene, but the twisting localizes the electrons and, ultimately, restores the order of the nanographene phase patterns. The electronic localization contributes to increase the HOMO-LUMO energy gap. When $N_t=6,7$, the restoring of the phase-patterns produces the sharp change noted in all plots displayed in Figure 2. At last, besides the HOMO and LUMO orbitals, we have also investigated the lowest energy $\pi$-like orbital for each $N_t$ and observed that, for even values of $N_t$, there is a continuous orbital ribbon on each face of the strip (even $N_t$ implies in an orientable surface) with no phase shifts, whereas for odd $N_t$ the corresponding orbital ribbons have at least one phase shift, as shown for $N_t=1,2$ in Figure 4.

According to the CI calculations, the singlet ground state and the first excited state -- which is a triplet for all nanoribbons and the nanographene -- are separated by an energy gap $E_{\rm{CI}}$ about 3 eV smaller (on average) than the HOMO-LUMO gap calculated using the AM1 Hamiltonian. Such energy gap reduction is due to the incorporation of some electronic correlation energy in the CI method. As $N_t$ is switched from 0 to 7, $E_{\rm{CI}}(N_t)$ varies qualitatively much like $E_g$. For the nanographene, $E_{\rm{CI}}=1.16$ eV. The 0-nanoribbon has $E_{\rm{CI}}=1.86$ eV and the half-twist M\"{o}bius band has $E_{\rm{CI}}=1.75$ eV. For the nanoribbons up to $N_t=6$ we have $E_{\rm{CI}}(N_t=2)=1.81$ eV, $E_{\rm{CI}}(N_t=3)=1.59$ eV,
$E_{\rm{CI}}(N_t=4)=1.63$ eV, $E_{\rm{CI}}(N_t=5)=1.38$ eV (the smallest of all energies), and $E_{\rm{CI}}(N_t=6)=1.93$ eV. It can be seen that $E_{\rm{CI}}(N_t)$ oscillates: an odd $N_t$ nanoribbon has a smaller energy gap than the value predicted for the even $N_t-1$ one. Exception to this rule is  $E_{\rm{CI}}(N_t=7)=2.03$ eV. Optical transitions between singlet and triplet states are forbidden by the spin selection rule, so the first excited state accessible through the absorption of photons has an energy larger than the first triplet excited quantum level. The plot at the bottom right of Figure 2 shows the first optical absorption transition energy ascribed to the nanoribbons,  $E_{\rm{OPT}}(N_t)$. It follows closely the plot for  $E_{\rm{CI}}$. The absorbed wavelengths corresponding to such transitions correspond to infrared (nanographene, $N_t=1,3,4,5$), orange ($N_t=0$), red ($N_t=2$), yellow ($N_t=6$), and green light ($N_t=7$). The intensity of the absorption peaks is related to the oscillator strength $f$, shown at the bottom of Figure 4. One can see that the nanographene has a very strong oscillator strength in comparison with the closed nanoribbons. $f(N_t)$ oscillates as we move from $N_t=0$ to $N_t=7$. For $N_t=0,2,4,6$ (even $N_t$) we have the minima of $f(N_t)$ whereas $N_t=1,3,5,7$ (odd $N_t$) correspond to maxima. The $N_t=0$ ribbon has the smallest value of $f$, $\approx 10^{-4}$ a.u., due to the symmetries of the ground state and first singlet excited state (indeed, one can say that this transition is practically forbidden). The first optical transition with significant oscillator strength for the 0-nanoribbon involves a singlet state with energy of 2.53 eV (corresponding to blue light), with $f \approx 9.4$ a.u. For $N_t=1$ (half-twist M\"{o}bius ribbon), the largest value of $f$ occurs for $E_{\rm{CI}}=2.43$ eV (blue light), and for $N_t=6,7$, $f$ is maximum for $E_{\rm{CI}}=2.86$ eV and $E_{\rm{CI}}=2.93$ eV (both equivalent to a blue wavelength), respectively. $f(N_t=5)$ ($\approx 0.8$ a.u.) is the largest of all first optically active electronic transitions shown in Figure 2.

Finally, at the top of Figure 5 the magnitude of the ground state dipole moment $p$ as a function of $N_t$ is shown. It was demonstrated that the use of configuration interaction including only HOMO and LUMO in the space of configurations leads to AM1+CI values for $p$ with practically the same quality as first principles results \cite{dipole}. The $N_t=3,6$ nanoribbons have the largest (and very close) dipole moments, $\approx 0.74$ D. In contrast, the $N_t=0$ ribbon has the smallest value, $\approx 10^{-2}$ D (5 times smaller than the dipole moment of the nanographene rectangle). In the case of a half-twist M\"{o}bius strip, $p(N_t)=0.32$ D. $p(N_t)$ has an order of alternate maxima and minima for $0\leq N_t \leq 5$ (minima for 0 and even $N_t$, maxima for odd $N_t$). This order is broken down for $N_t=6$. We think that crystals made from some of the nanoribbons (specifically those with $N_t$ odd and the $N_t=6$ case) could be stabilized through dipole-dipole interactions, as it occurs with many organic compounds.

\section{Conclusions}

In this paper, we presented the results of computational simulations to obtain structural and electronic properties of twisted nanoribbons built from a nanographene strip. Optimized geometries were found through classical molecular mechanics calculations in three steps: (i) first geometry optimization starting from symmetric initial structures; (ii) thermal annealing to search for geometries with lower total energy; (iii) second geometry optimization of the nanostructures with lowest total energy found after the annealing. A stability study was performed on the optimal geometries to check for the dissociation of chemical bonds. All nanoribbons were stable for a temperature of 500 K and simulation time of 1 ns. The total energy as a function of the number of twists was fitted to a parabolic curve with good accuracy, the largest fit error being observed for the $N_t=1$ (half-twist M\"{o}bius band) case. Optimized geometries have lower symmetry in comparison with starting configurations.

The electronic structure was calculated for all nanoribbons using the AM1 semiempirical Hamiltonian and a CI calculation was carried out to improve the estimate of the HOMO-LUMO energy gap, which is larger for the twisted molecules in comparison to the flat metallic nanographene. We noted a sharp increase of the HOMO$\rightarrow$LUMO transition energy as the number of twists varies from $N_t=5$ to $N_t=6,7$, indicating that there is a change of electronic structure related to the increase of torsion in the twisted nanoribbons. When a closed nanoribbon ($N_t=0$) is formed, the HOMO-LUMO wavefunction appearance is inverted in comparison to the armchair nanographene case. Increasing $N_t$ leads to electronic localization in the frontier orbitals and restores the nanoribbon HOMO-LUMO patterns to those of the nanographene. AM1+CI energy gaps are much smaller than the ones calculated using the AM1 Hamiltonian, but probably much larger than the DFT-GGA estimates. The nanographene oscillator strength $f$ for the first optically active electronic transition is stronger than the corresponding ones for the closed nanoribbons. For $N_t=0,2,4,6$ (even $N_t$), $f(N_t)$ has minima, whereas $N_t=1,3,5,7$ (odd $N_t$) correspond to maxima of $f$. The first allowed optical transition for the case $N_t=0$ between the singlet ground state and the singlet excited state is forbidden due to the wavefunction symmetries. For $N_t=1$ (half-twist M\"{o}bius nanoribbon), $f$ is maximum for $E_{\rm{CI}}=2.43$ eV. Crystals made from the nanoribbons with $N_t$ odd and $N_t=6$ could possibly be stabilized through dipolar interactions.

\textbf{Acknowledgements}

E. W. S. Caetano thanks the support received from the Brazilian
National Research Council (Conselho Nacional de Pesquisa - CNPq) through the process 478885/2006-7 Edital MCT/CNPq 02/2006 - Universal and from the CEFET-CE/ProAPP research program.

\clearpage

\renewcommand{\baselinestretch}{1}

\begin{table}[h]
\begin{center}
\caption{Space point groups of the graphene nanoribbons studied in this work: initial geometry (second column), classically optimized geometry (after annealing, third column), and semiempirical optimized geometry (fourth column). The second row shows the space point group of the nanographene rectangle used to build the nanoribbons.} \vskip1truecm
\begin{tabular}{llll}
\hline
\multicolumn{1}{c}{Number of} & \multicolumn{1}{c}{Initial} & \multicolumn{1}{c}{Optimized} & \multicolumn{1}{c}{Optimized} \\
\multicolumn{1}{c}{twists ($N_t$)} & \multicolumn{1}{c}{geometry} & \multicolumn{1}{c}{(annealing)} & \multicolumn{1}{c}{(semiempirical)} \\
\hline
\multicolumn{1}{c}{Nanographene} & \multicolumn{1}{c}{$D_{2h}$} & \multicolumn{1}{c}{-} & \multicolumn{1}{c}{-} \\
\multicolumn{1}{c}{0} & \multicolumn{1}{c}{$D_{15h}$} & \multicolumn{1}{c}{$C_2$} & \multicolumn{1}{c}{$C_{5h}$} \\
\multicolumn{1}{c}{1} & \multicolumn{1}{c}{$C_2$} & \multicolumn{1}{c}{$C_1$} & \multicolumn{1}{c}{$C_1$} \\
\multicolumn{1}{c}{2} & \multicolumn{1}{c}{$C_2$} & \multicolumn{1}{c}{$C_1$} & \multicolumn{1}{c}{$C_1$} \\
\multicolumn{1}{c}{3} & \multicolumn{1}{c}{$C_3$} & \multicolumn{1}{c}{$C_2$} & \multicolumn{1}{c}{$C_1$} \\
\multicolumn{1}{c}{4} & \multicolumn{1}{c}{$C_2$} & \multicolumn{1}{c}{$C_1$} & \multicolumn{1}{c}{$C_1$} \\
\multicolumn{1}{c}{5} & \multicolumn{1}{c}{$C_5$} & \multicolumn{1}{c}{$C_2$} & \multicolumn{1}{c}{$C_2$} \\
\multicolumn{1}{c}{6} & \multicolumn{1}{c}{$C_3$} & \multicolumn{1}{c}{$C_2$} & \multicolumn{1}{c}{$C_1$} \\
\multicolumn{1}{c}{7} & \multicolumn{1}{c}{$C_2$} & \multicolumn{1}{c}{$C_1$} & \multicolumn{1}{c}{$C_1$} \\
\hline
\end{tabular}
\end{center}
\end{table}

\clearpage

\begin{table}[h]
\begin{center}
\caption{Total energy and its components (bond, angle, torsion, inversion, and van der Waals) for the optimized molecules after the classical annealing. The percentual of total energy for each contribution is also shown. All energies are given in kcal/mol.} \vskip1truecm
\begin{tabular}{lllllll}
\hline
\multicolumn{1}{c}{} & \multicolumn{1}{c}{Total} & \multicolumn{1}{c}{Bond} & \multicolumn{1}{c}{Angle} & \multicolumn{1}{c}{Torsion} & \multicolumn{1}{c}{Inversion} & \multicolumn{1}{c}{vdW} \\
\multicolumn{1}{c}{$N_t$} & \multicolumn{1}{c}{Energy} & \multicolumn{1}{c}{Energy} & \multicolumn{1}{c}{Energy} & \multicolumn{1}{c}{Energy} & \multicolumn{1}{c}{Energy} & \multicolumn{1}{c}{Energy} \\
\hline
\multicolumn{1}{c}{0} & \multicolumn{1}{c}{296.74} & \multicolumn{1}{c}{31.97} & \multicolumn{1}{c}{16.99} & \multicolumn{1}{c}{63.22} & \multicolumn{1}{c}{1.79} & \multicolumn{1}{c}{182.77} \\
\multicolumn{1}{c}{1} & \multicolumn{1}{c}{321.65} & \multicolumn{1}{c}{30.18} & \multicolumn{1}{c}{18.31} & \multicolumn{1}{c}{91.78} & \multicolumn{1}{c}{3.57} & \multicolumn{1}{c}{177.81} \\
\multicolumn{1}{c}{2} & \multicolumn{1}{c}{331.70} & \multicolumn{1}{c}{29.55} & \multicolumn{1}{c}{18.00} & \multicolumn{1}{c}{103.58} & \multicolumn{1}{c}{4.47} & \multicolumn{1}{c}{176.10} \\
\multicolumn{1}{c}{3} & \multicolumn{1}{c}{372.52} & \multicolumn{1}{c}{28.97} & \multicolumn{1}{c}{22.14} & \multicolumn{1}{c}{158.38} & \multicolumn{1}{c}{10.67} & \multicolumn{1}{c}{152.36} \\
\multicolumn{1}{c}{4} & \multicolumn{1}{c}{423.25} & \multicolumn{1}{c}{29.32} & \multicolumn{1}{c}{30.42} & \multicolumn{1}{c}{207.13} & \multicolumn{1}{c}{15.90} & \multicolumn{1}{c}{140.48} \\
\multicolumn{1}{c}{5} & \multicolumn{1}{c}{501.19} & \multicolumn{1}{c}{30.74} & \multicolumn{1}{c}{42.80} & \multicolumn{1}{c}{273.71} & \multicolumn{1}{c}{24.32} & \multicolumn{1}{c}{129.62} \\
\multicolumn{1}{c}{6} & \multicolumn{1}{c}{613.12} & \multicolumn{1}{c}{33.58} & \multicolumn{1}{c}{54.99} & \multicolumn{1}{c}{351.98} & \multicolumn{1}{c}{35.50} & \multicolumn{1}{c}{137.07} \\
\multicolumn{1}{c}{7} & \multicolumn{1}{c}{770.37} & \multicolumn{1}{c}{37.66} & \multicolumn{1}{c}{96.03} & \multicolumn{1}{c}{441.31} & \multicolumn{1}{c}{53.18} & \multicolumn{1}{c}{142.19} \\
\hline
\multicolumn{1}{c}{} & \multicolumn{1}{c}{} & \multicolumn{1}{c}{Bond/} & \multicolumn{1}{c}{Angle/} & \multicolumn{1}{c}{Torsion/} & \multicolumn{1}{c}{Inversion/} & \multicolumn{1}{c}{vdW/} \\
\multicolumn{1}{c}{$N_t$} & \multicolumn{1}{c}{} & \multicolumn{1}{c}{Total (\%)} & \multicolumn{1}{c}{Total (\%)} & \multicolumn{1}{c}{Total (\%)} & \multicolumn{1}{c}{Total (\%)} & \multicolumn{1}{c}{Total (\%)} \\
\hline
\multicolumn{1}{c}{0} & \multicolumn{1}{c}{} & \multicolumn{1}{c}{10.77} & \multicolumn{1}{c}{5.73} & \multicolumn{1}{c}{21.30} & \multicolumn{1}{c}{0.60} & \multicolumn{1}{c}{61.60} \\
\multicolumn{1}{c}{1} & \multicolumn{1}{c}{} & \multicolumn{1}{c}{9.38} & \multicolumn{1}{c}{5.69} & \multicolumn{1}{c}{28.53} & \multicolumn{1}{c}{1.11} & \multicolumn{1}{c}{55.29} \\
\multicolumn{1}{c}{2} & \multicolumn{1}{c}{} & \multicolumn{1}{c}{8.91} & \multicolumn{1}{c}{5.43} & \multicolumn{1}{c}{31.23} & \multicolumn{1}{c}{1.35} & \multicolumn{1}{c}{53.08} \\
\multicolumn{1}{c}{3} & \multicolumn{1}{c}{} & \multicolumn{1}{c}{7.78} & \multicolumn{1}{c}{5.94} & \multicolumn{1}{c}{42.52} & \multicolumn{1}{c}{2.86} & \multicolumn{1}{c}{40.90} \\
\multicolumn{1}{c}{4} & \multicolumn{1}{c}{} & \multicolumn{1}{c}{6.93} & \multicolumn{1}{c}{7.19} & \multicolumn{1}{c}{48.94} & \multicolumn{1}{c}{3.76} & \multicolumn{1}{c}{33.18} \\
\multicolumn{1}{c}{5} & \multicolumn{1}{c}{} & \multicolumn{1}{c}{6.13} & \multicolumn{1}{c}{8.54} & \multicolumn{1}{c}{54.61} & \multicolumn{1}{c}{4.85} & \multicolumn{1}{c}{25.87} \\
\multicolumn{1}{c}{6} & \multicolumn{1}{c}{} & \multicolumn{1}{c}{5.48} & \multicolumn{1}{c}{8.97} & \multicolumn{1}{c}{57.41} & \multicolumn{1}{c}{5.79} & \multicolumn{1}{c}{22.35} \\
\multicolumn{1}{c}{7} & \multicolumn{1}{c}{} & \multicolumn{1}{c}{4.89} & \multicolumn{1}{c}{12.47} & \multicolumn{1}{c}{57.29} & \multicolumn{1}{c}{6.90} & \multicolumn{1}{c}{18.45} \\
\hline
\end{tabular}
\end{center}
\end{table}

\clearpage

\begin{figure}[t]
\centerline{\includegraphics[width=0.70\textwidth]{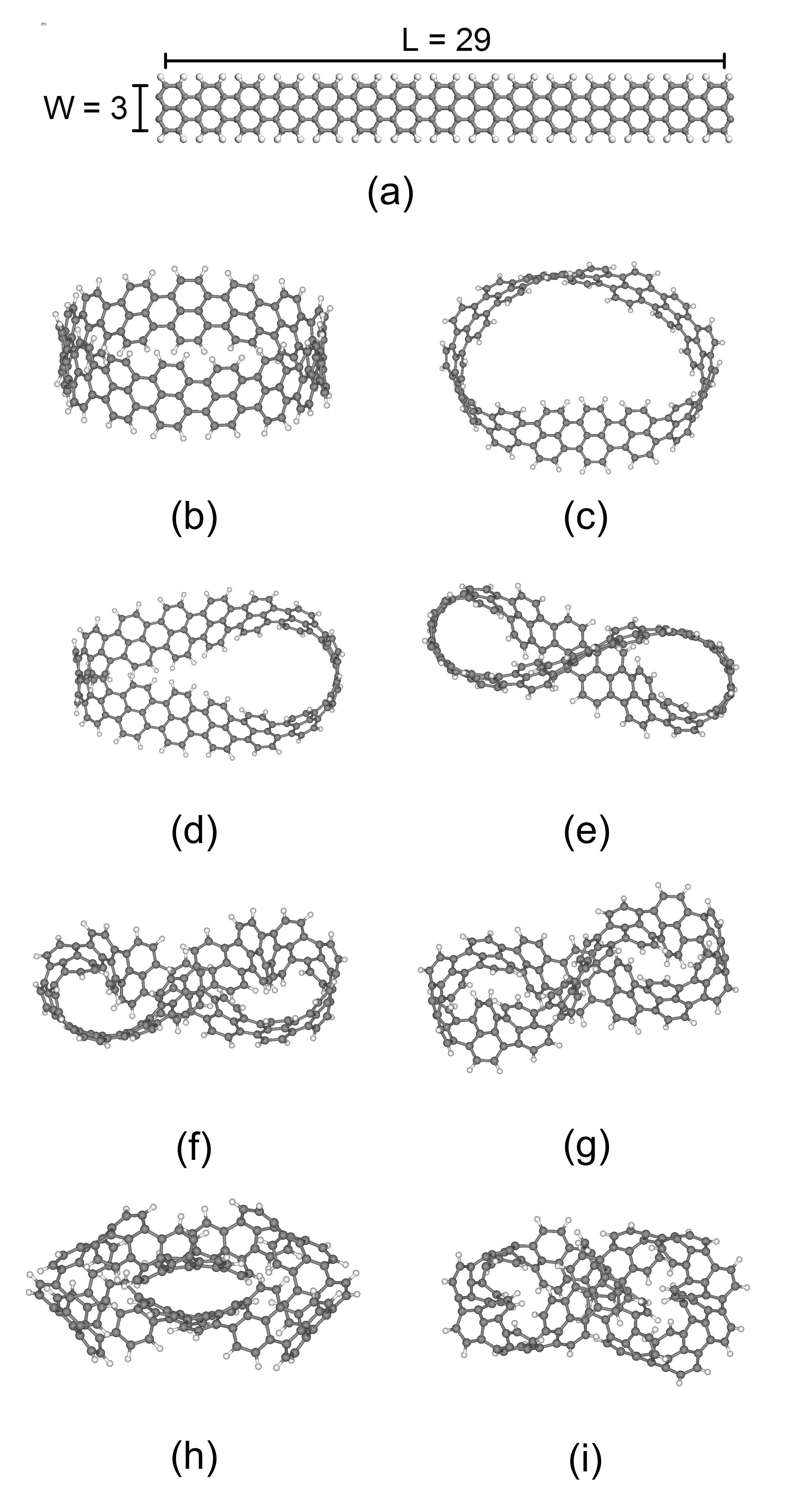}}
\caption{(a) Rectangular nanographene used to build the twisted nanoribbons. The twisted nanoribbons, after classical annealing and geometry optimization, are presented; (b) N$_t=0$ nanoribbon; (c) N$_t=1$ nanoribbon (half-twist M\"{o}bius strip); (d) N$_t=2$; (e) N$_t=3$; (f) N$_t=4$; (g) N$_t=5$; (h) N$_t=6$; (i) N$_t=7$.} \label{figure1}
\end{figure}

\clearpage

\begin{figure}[t]
\centerline{\includegraphics[width=1.0\textwidth]{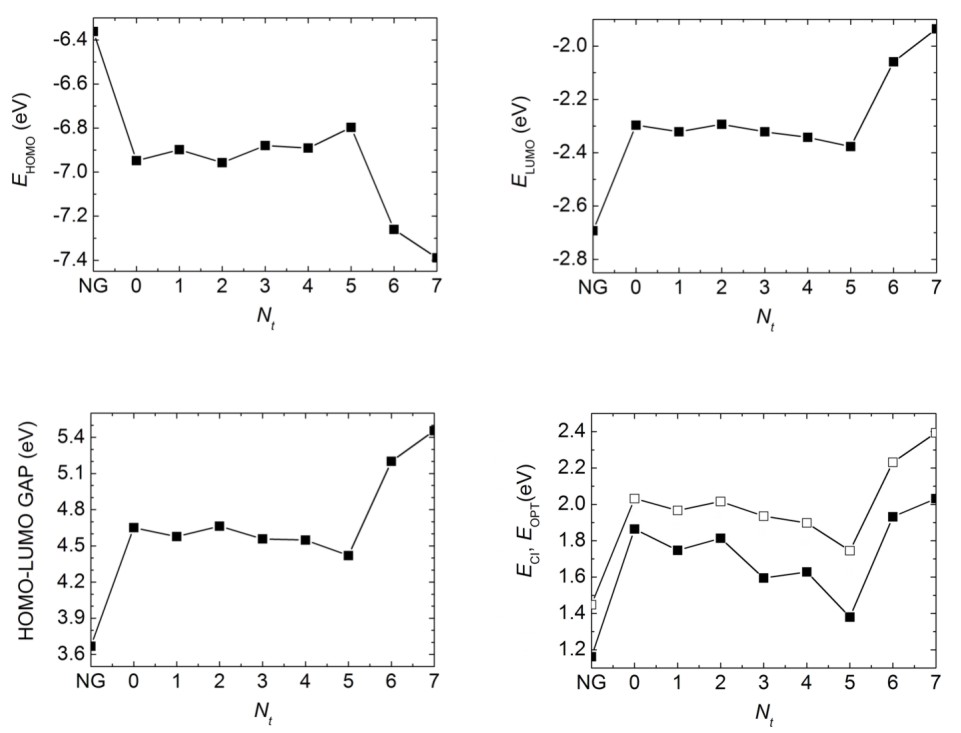}}
\caption{HOMO and LUMO energy levels, HOMO-LUMO energy gap, CI energy gap ($E_{\rm{CI}}$, solid squares) and the energy of the first allowed optical transitions ($E_{\rm{OPT}}$, open squares) for all the nanoribbons and the nanographene rectangle (NG).} \label{figure2}
\end{figure}

\clearpage

\begin{figure}[t]
\centerline{\includegraphics[width=1.0\textwidth]{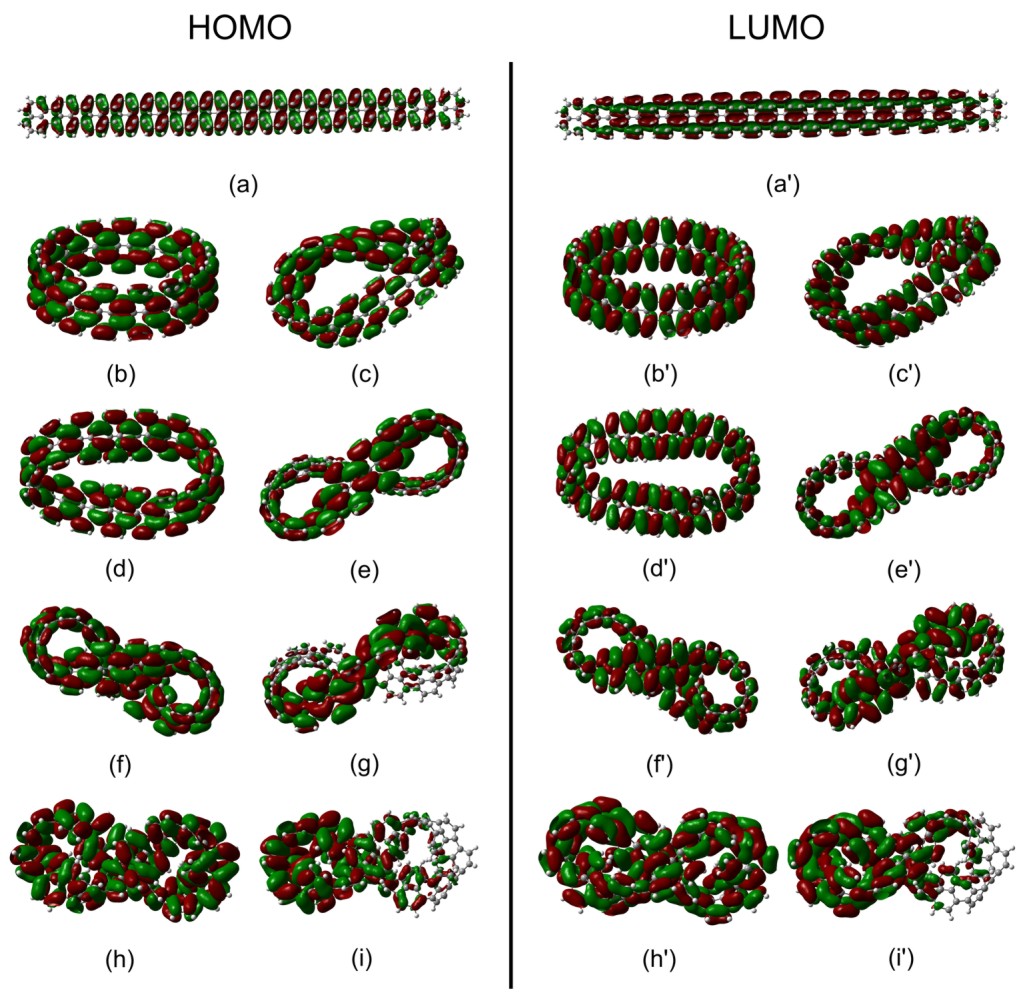}}
\caption{HOMO and LUMO orbitals for the: (a,a') Rectangular nanographene; (b,b') N$_t=0$ nanoribbon; (c,c') N$_t=1$ nanoribbon (half-twist M\"{o}bius strip); (d,d') N$_t=2$; (e,e') N$_t=3$; (f,f') N$_t=4$; (g,g') N$_t=5$; (h,h') N$_t=6$; (i,i') N$_t=7$. Isosurfaces correspond to a wavefunction amplitude of 0.002.} \label{figure3}
\end{figure}

\clearpage

\begin{figure}[t]
\centerline{\includegraphics[width=1.0\textwidth]{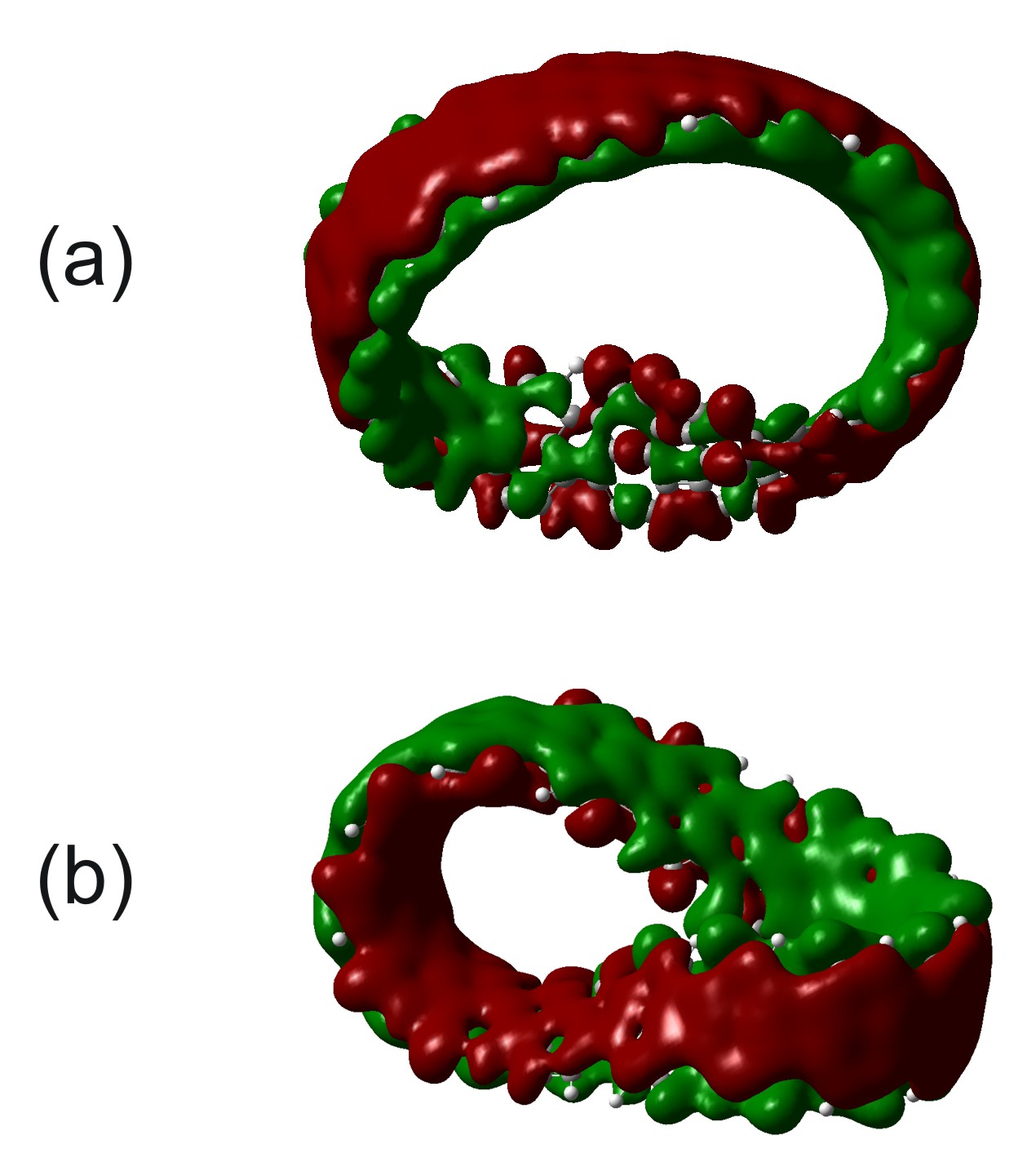}}
\caption{Lowest energy $\pi$-like orbitals for $N_t=1$ (a) and $N_t=2$ nanoribbons. For the $N_t=1$ case, we note at the localized twist the appearance of a phase shift in the wavefunction. There is no phase shift in the $N_t=2$ ribbon.} \label{figure4}
\end{figure}

\clearpage

\begin{figure}[t]
\centerline{\includegraphics[width=0.8\textwidth]{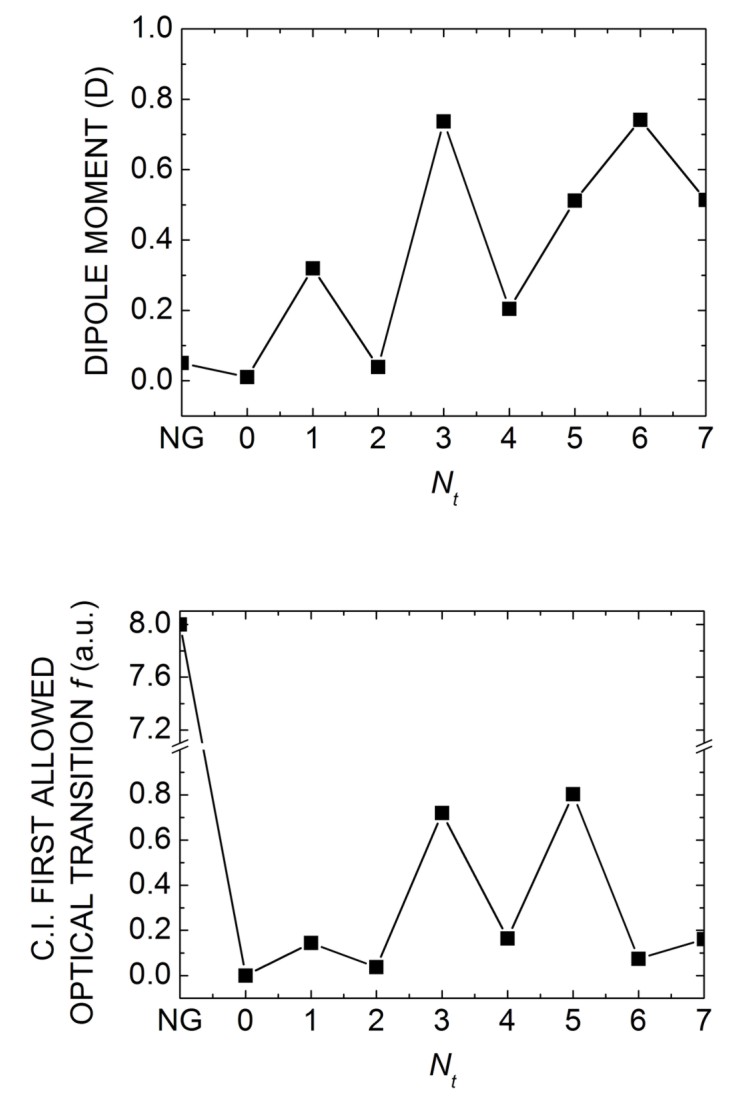}}
\caption{Dipole moment and oscillator strength of the first allowed optical transition for each graphene nanoribbon and for the nanographene molecule (NG).} \label{figure5}
\end{figure}

\end{document}